\documentclass[twocolumn,secnumarabic,amssymb, nobibnotes, aps, prd]{revtex4-1}

\usepackage{graphicx}

\usepackage[usenames]{color}

\newcommand{\ttbar}     {\mbox{$t\bar{t}$}}

\newcommand{\ppbar}     {\mbox{$p\bar{p}$}}

\def\bea{\begin{eqnarray}}
\def\eea{\end{eqnarray}}
\def\half{\frac{1}{2}}

\begin{document}

\title{Cross Sections for Leptophobic Topcolor $Z'$ decaying to top-antitop}%

\begin{flushright}
FERMILAB-FN-0940-CMS-PPD \\
\end{flushright}

\author{{
Robert M.~Harris,$^{1}$
Supriya Jain$^{2}$
}}

\affiliation{\large{\vspace*{0.1in}
$^{1}$Fermi National Accelerator Laboratory \\
$^{2}$State University of New York at Buffalo 
}}

\date{June 19, 2012}%

\begin{abstract}
{{We present numerical calculations of the production cross section 
    of a heavy $Z'$ resonance in hadron-hadron collisions with subsequent 
    decay into top-antitop pairs. In particular, we consider the leptophobic 
    topcolor $Z'$ discussed under Model IV of hep-ph/9911288 which has  
    predicted cross sections large enough to be experimentally accessible 
    at the Fermilab Tevatron and the Large Hadron Collider at CERN. 
    This article presents an updated calculation valid for the
    Tevatron and all proposed LHC collision energies.
    Cross sections are presented for various $Z'$ widths, in $\ppbar$
    collisions at $\sqrt{s}=2$~TeV, and in $pp$ collisions at 
    $\sqrt{s}=7$, 8, 10 and 14~TeV.
}}
\end{abstract}
\pacs{12.20.Ds, 12.60.Cn, 12.60.Nz, 14.65.Ha, 14.70.Pw, 14.80.Tt}
%
%

\maketitle
\tableofcontents


\section{Introduction}
\label{sec:introduction}

Electroweak symmetry breaking is a cornerstone for our understanding of 
particle physics. However, despite the spectacular phenomenological success 
of the Standard Model (SM), the fundamental mechanism of electroweak 
symmetry breaking remains a mystery. Various new models have been proposed 
to explain this mechanism. One such class of models is based upon 
topcolor~\cite{ref_topc1,ref_topc2} which can generate a large
top-quark mass. The topcolor model also predicts 
a $Z'$~\cite{hep-ph-9911288}.  

The physics in production and decay of the $Z'$ are discussed in 
Ref.~\cite{hep-ph-9911288} under different model assumptions. In this 
paper, we consider the 
$Z'$ from Model IV which represents a novel class and has predicted
cross sections large enough to be experimentally accessible at hadron
colliders at the Fermilab Tevatron and the Large Hadron Collider (LHC) at
CERN. Such $Z'$ resonances couple strongly only to the first and third 
generations of quarks, and have no significant couplings to the
leptons. They are, therefore, leptophobic and topophyllic.

This article presents an updated calculation valid for the Tevatron
and all proposed LHC collision energies. The leptophobic 
topcolor $Z'$ decaying to $t\bar{t}$ has
been searched for at both the Tevatron~\cite{tev1,tev2,tev3} and the 
LHC~\cite{lhc1}. The Tevatron searches used a previous calculation of the
$Z'$ cross section~\cite{hep-ph-9911288}. The LHC search conducted by 
the CMS collaboration have used the cross section calculation presented
here to set limits on the $Z'$ mass. This article is primarily
intended to document that $Z'$ cross section used for a $pp$ collision
energy of $\sqrt{s} = 7$~TeV.
For current and future LHC searches, we present the $Z'$ cross section
at the current LHC $pp$ collision energy of $\sqrt{s} = 8$~TeV and the potential
future collision energies of 10 and 14~TeV.  We also present
the cross section in $p\bar{p}$ collisions at $\sqrt{s} = 2$~TeV for
comparison with Tevatron searches.

\section{Model}
\label{sec:model}
As discussed in Ref.~\cite{hep-ph-9911288}, non-standard models 
can be constructed in which the 
$U(1)_Y\rightarrow U(1)_1\times U(1)_2 $
and the generations are grouped differently. 
Model IV is the case where quark generations 
$(1,3) \supset U(1)_2$. The dominant part of the 
interaction Lagrangian for Model IV is:
\bea
L_{IV} & = &
(\half g_1\cot\theta_H)Z'^{\mu}\left(  
\bar{t}_L\gamma_\mu t_L 
+\bar{b}_L\gamma_\mu b_L
+ f_1\bar{t}_R\gamma_\mu  t_R 
\right. \nonumber \\
& &  \left.
+ f_2\bar{b}_R\gamma_\mu  b_R 
-\bar{u}_L\gamma_\mu u_L 
-\bar{d}_L\gamma_\mu d_L
-f_1\bar{u}_R\gamma_\mu  u_R 
\right. \nonumber \\
& &  \left.
-f_2\bar{d}_R\gamma_\mu  d_R 
\right),
\eea
\noindent
where $g_1$ is the SM coupling constant, and we require the 
following: 
$f_1 > 0$ (attractive $\bar{t}t$ channel)
and/or $f_2 <0 $ (repulsive $\bar{b}b$ channel).
Also, $\cot\theta_H >> 1$ to avoid fine-tuning. 
The parton-level subprocess cross
section that we use in the next section is obtained from this
Lagrangian. 

\section{Cross Section for $Z'$ Production and Decay}
\label{sec:xsecFormulae}

The total lowest-order cross section for a $Z'$ produced in hadron
collisions and decaying into top-antitop pairs from Model IV discussed 
above, is given by:
\begin{equation}
\small{
\sigma_{Z'} \times {\rm B(}Z'\rightarrow\ttbar{\rm )} \equiv 
\sigma = \int_{0}^{\infty} \frac{d\sigma}{dm} dm
}
\label{eq_tot_xsec}
\end{equation}
where $d\sigma/dm$, the differential cross section at 
$t\bar{t}$ invariant mass $m$, is given by 
\begin{equation}
\small{
\frac{d\sigma}{dm} = \frac{2}{m} \int_{-\ln(\sqrt{s}/m)}^{\ln(\sqrt{s}/m)} dy_b \ \tau 
{\cal L}(x_{h1},x_{h2}) \ \hat{\sigma}(q\bar{q} \rightarrow Z' \rightarrow t\bar{t}). 
}
\label{eq_xsec}
\end{equation}
Here $\hat{\sigma}(q\bar{q} \rightarrow Z' \rightarrow t\bar{t})$ 
is the parton-level subprocess cross section. The kinematic variable
$\tau$ is related to the initial-state parton fractional momenta inside
the first hadron $x_{h1}$ and the second hadron $x_{h2}$ by 
$\tau = x_{h1} x_{h2} = m^2/s$.  The boost of the partonic system 
$y_b$ is given by $y_b = (1/2) \ln(x_{h1}/x_{h2})$.
The partonic ``luminosity function'' is just the product of parton
distribution functions:
\begin{equation}
\small{
{\cal L}(x_{h1},x_{h2}) = q(x_{h1},\mu)\bar{q}(x_{h2},\mu) + \bar{q}(x_{h1},\mu)q(x_{h2},\mu)
}
\end{equation}
where $q(x, \mu)$ ($\bar{q}(x,\mu)$) is the parton distribution function of a 
quark (anti-quark) evaluated at fractional momenta $x$ and renormalization 
scale $\mu$.

The parton-level subprocess cross section, 
$\hat{\sigma}(q\bar{q} \rightarrow Z' \rightarrow t\bar{t})$, for 
a leptophobic, b$_r$-phobic, and topophyllic, $Z'$ is given by: 
\begin{widetext}
\begin{eqnarray}
\hat\sigma
& \rightarrow &
\frac{9 \alpha^2 \pi  }{16\cos^4\theta_W}\cot^4\theta_H
\times \left(2 \; \makebox{for initial
state $u+\bar{u}$;}, 
\; (1) \;\makebox{for initial $d+\bar{d}$ }\right)
\nonumber \\ 
& & \times
\left[ 2 \times \beta(1 +\frac{1}{3}\beta^2)
+ \beta (1-\beta^2) \right] 
\left[ \frac{s}{(\hat{s}-M_{Z'}^2)^2 + \hat{s}\Gamma_{Z'}^2 } \right]\theta(\hat{s}-4m_t^2),
\label{eq_sigma_IV}
\end{eqnarray}
\end{widetext}
where $\cot^4\theta_H$ can be obtained from the total decay-width of
the resonance~\cite{hep-ph-9911288-update}:
\begin{equation}
\small{
\Gamma_{Z'} = 
\frac{ \alpha\cot^2\theta_H M_{Z'}}{8 \cos^2\theta_W } 
\left[ \sqrt{ 1 - \frac{4m_t^2}{M_{Z'}^2} } 
\left( 2 + 4\frac{m_t^2}{M_{Z'}^2} 
\right) 
+ 4
\right].
}
\label{eq_gamma_IV}
\end{equation}
The subprocess cross section in 
Eq.~\ref{eq_sigma_IV} is for spin and color
summing on both initial and final state legs, while most parton distributions
assume spin and color averaged on the initial-state legs and spin and color
summing on the final-state legs.  Therefore, the subprocess cross section 
given by Eq.~\ref{eq_sigma_IV} must be multiplied by a factor of
\begin{equation}
\small{
\left(\frac{1}{spins}\right)^2\left(\frac{1}{colors}\right)^2 = 
\left(\frac{1}{2}\right)^2\left(\frac{1}{3}\right)^2 = \frac{1}{36}
}
\label{eq_spin_color}
\end{equation}
when used with parton distributions from PDFLIB~\cite{ref_pdflib} and 
other standard sources. We have taken this into account when
calculating the cross section.
We have also used $m_t=172.5$ GeV/c$^2$, and $\cos^2\theta_W=0.768$. 
For a default parton distribution set we have chosen CTEQ6L~\cite{ref_cteq}. 
This is a modern parton distribution set appropriate for leading-order
calculations and is available in PDFLIB~\cite{ref_pdflib}. For a
default renormalization scale we choose
$\mu=m/2$, half the $t\bar{t}$  invariant mass.  This scale has the
benefit that it reduces to the usual $\mu=m_t$ at top production threshold,
but also increases with increasing $t\bar{t}$ invariant mass. With these
choices, the total cross section for the production of a leptophobic
and topophyllic $Z'$ and its subsequent decay into $\ttbar$ pairs is
presented in Sections~\ref{sec:tevatron}~and~\ref{sec:lhc} for the
Tevatron and the LHC. 


\section{Cross sections at the Tevatron}
\label{sec:tevatron}
\begin{figure}[!htb]
 \centering
  \includegraphics[width=0.5\textwidth,angle=0]{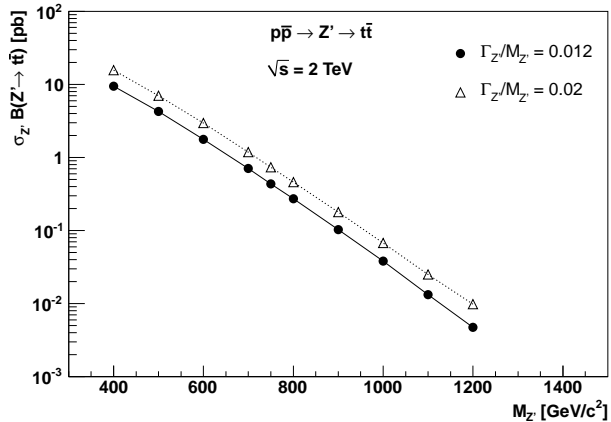}
\caption{Cross sections at the Tevatron at $\sqrt{s} = 2$~TeV, for 
$\sigma_{Z'}{\rm B(}Z'\rightarrow\ttbar{\rm )}$, with
different choices of the resonance width.
\label{fig:tevatron}}
\end{figure}

We calculate numerically the lowest order cross section for the process 
$\ppbar \rightarrow Z' \rightarrow \ttbar$ at the Tevatron 
at $\sqrt{s} = 2$~TeV, using Eq.~\ref{eq_tot_xsec} and taking into account 
the spin-color factor in Eq.~\ref{eq_spin_color}. 
The only remaining parameter of the topcolor
model that affects the cross section is the mixing angle
$\cot^2\theta_H$, or equivalently the width $\Gamma_{Z'}$ which is
related to it. We calculate the cross section for different choices of 
$\Gamma_{Z'}$, equal to $1.2\%$ and $2\%$ of $M_{Z'}$, both of which
qualify as narrow resonances at the Tevatron. The integration
in Eq.~\ref{eq_tot_xsec} is performed using the full available
phase space of $ 2m_t < m < \sqrt{s}$. The results are tabulated in
Table~\ref{tab:ZprimexsecTev2} and displayed in Fig.~\ref{fig:tevatron}. 
\begin{table}[htbH]
\begin{center}
\caption{Cross sections at the Tevatron at $\sqrt{s} = 2$~TeV for 
$\sigma_{Z'}{\rm B(}Z'\rightarrow\ttbar{\rm )}$ at different $Z'$ 
masses and widths.
\label{tab:ZprimexsecTev2}}
\begin{ruledtabular}
\begin{tabular}{c|cc}
$M_{Z'}$ & \multicolumn{2}{c}{$\sigma_{Z'}{\rm B(}Z'\rightarrow\ttbar{\rm )}$
  [pb]} \\ \cline{2-3}
[GeV/c$^2$] 
& $\Gamma_{Z'}/M_{Z'} = 0.012$ 
& $\Gamma_{Z'}/M_{Z'} = 0.02$ \\
\hline 
400.0	&	9.49	&	15.62	\\
500.0	&	4.25	&	7.03	\\
600.0	&	1.77	&	2.95	\\
700.0	&	0.705	&	1.18	\\
750.0	&	0.435	&	0.735	\\
800.0	&	0.273	&	0.462	\\
900.0	&	0.103	&	0.178	\\
1000.0	&	3.80E-02	&	6.76E-02	\\
1100.0	&	1.33E-02	&	2.50E-02	\\
1200.0	&	4.75E-03	&	9.81E-03	\\
\end{tabular}
\end{ruledtabular}
\end{center}
\end{table}
%


\section{Cross sections at the LHC}
\label{sec:lhc}
\begin{figure}\tt
 \centering
  \includegraphics[width=0.5\textwidth,angle=0]{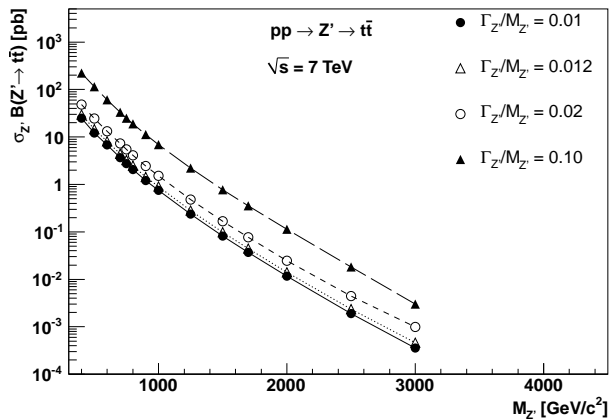}
\caption{Cross sections at the LHC at $\sqrt{s} = 7$~TeV, for 
$\sigma_{Z'}{\rm B(}Z'\rightarrow\ttbar{\rm )}$, with
different choices of the resonance width.
\label{fig:LHC7TeV}}
\end{figure}

\begin{figure}\tt
 \centering
  \includegraphics[width=0.5\textwidth,angle=0]{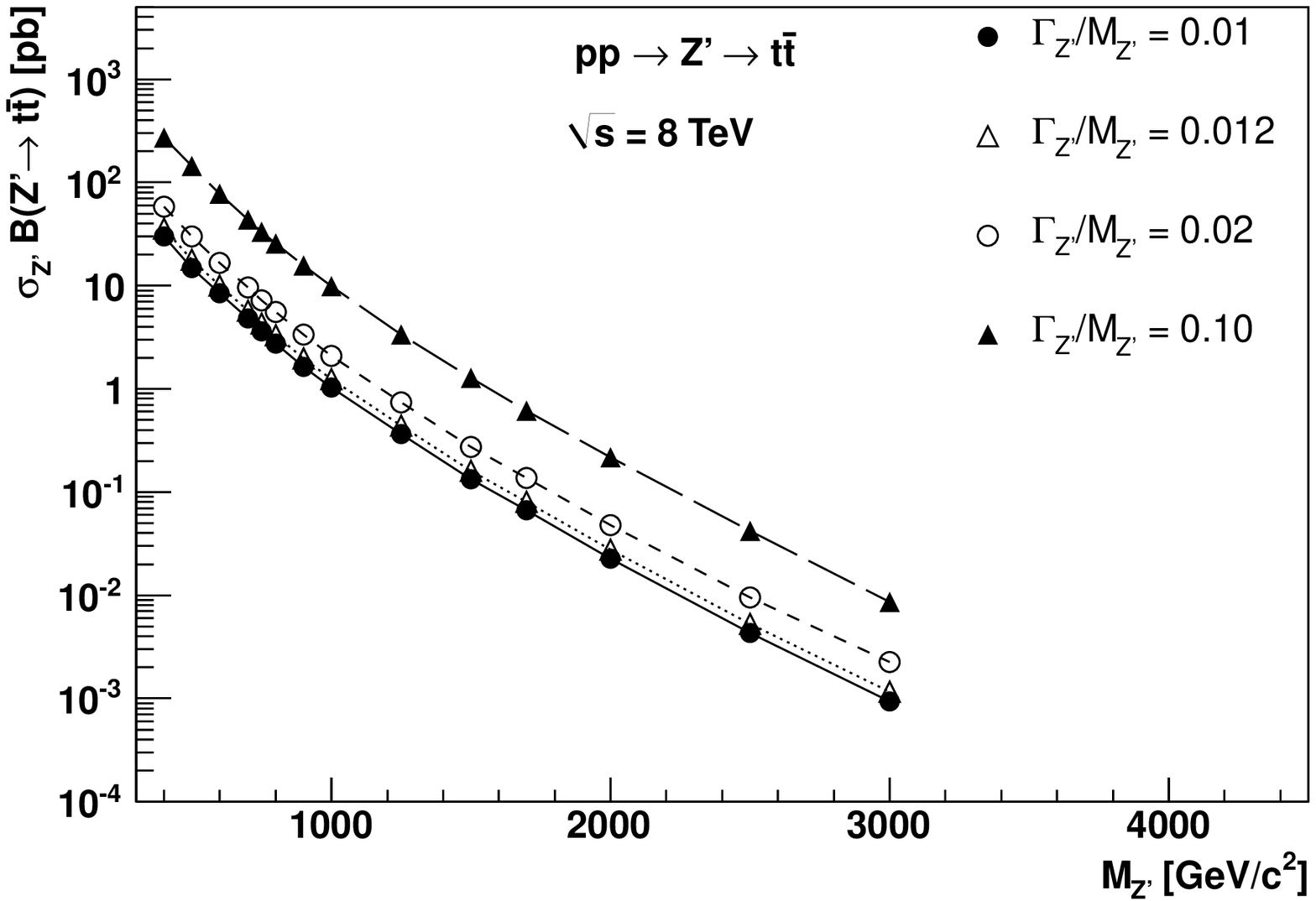}
\caption{Cross sections at the LHC at $\sqrt{s} = 8$~TeV, for 
$\sigma_{Z'}{\rm B(}Z'\rightarrow\ttbar{\rm )}$, with
different choices of the resonance width.
\label{fig:LHC8TeV}}
\end{figure}

\begin{figure}\tt
 \centering
  \includegraphics[width=0.5\textwidth,angle=0]{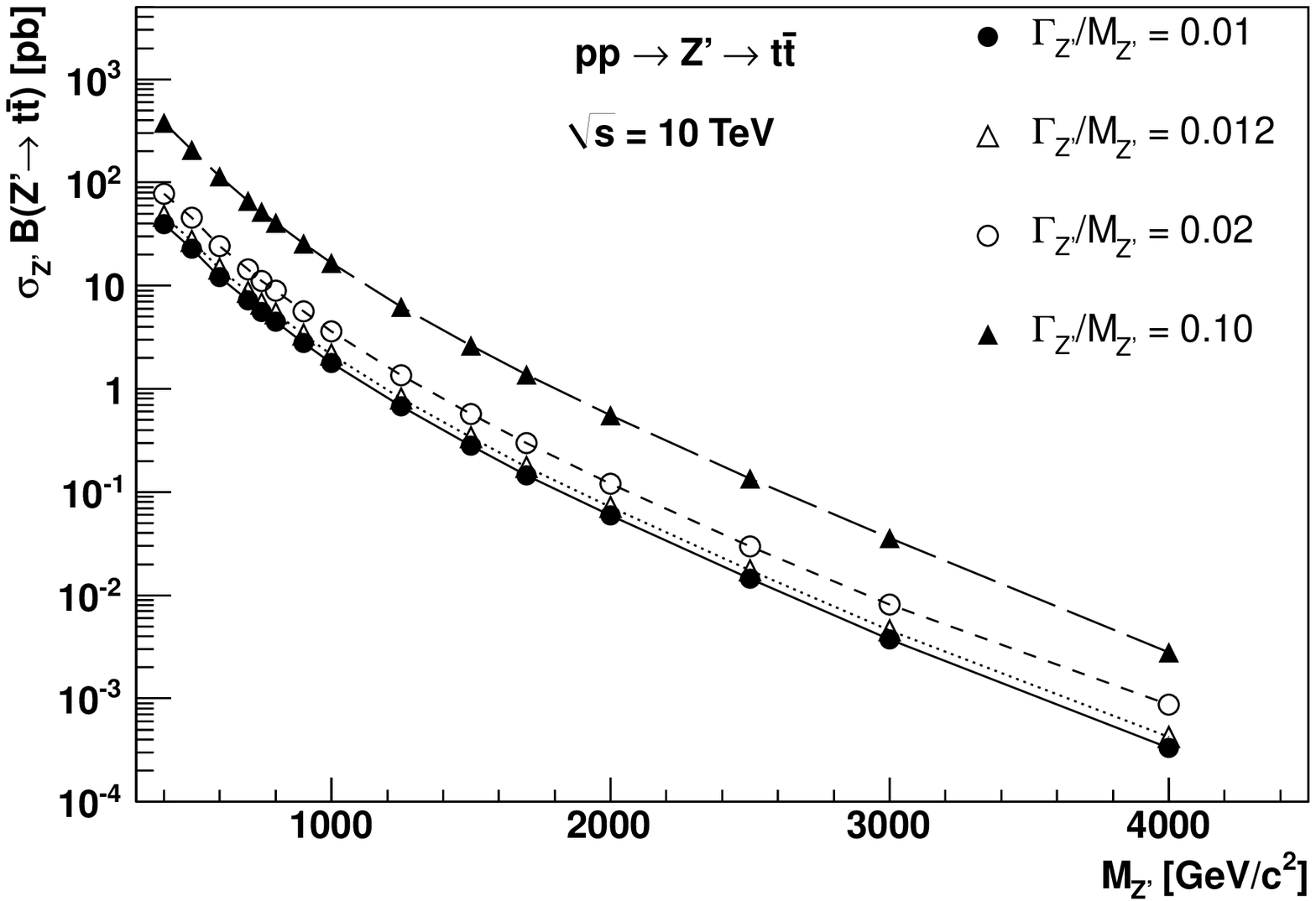}
\caption{Cross sections at the LHC at $\sqrt{s} = 10$~TeV, for 
$\sigma_{Z'}{\rm B(}Z'\rightarrow\ttbar{\rm )}$, with
different choices of the resonance width.
\label{fig:LHC10TeV}}
\end{figure}

\begin{figure}\tt
 \centering
  \includegraphics[width=0.5\textwidth,angle=0]{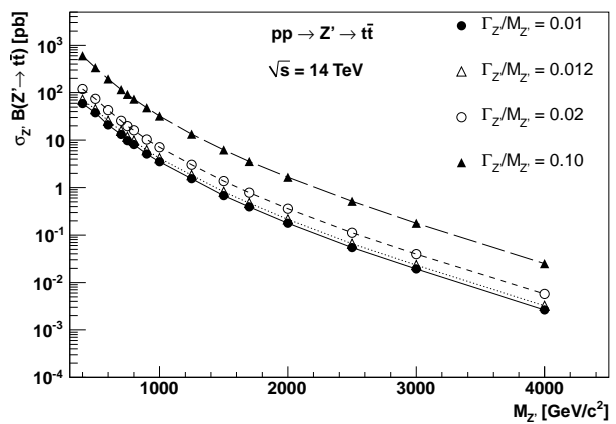}
\caption{Cross sections at the LHC at $\sqrt{s} = 14$~TeV, for 
$\sigma_{Z'}{\rm B(}Z'\rightarrow\ttbar{\rm )}$, with
different choices of the resonance width.
\label{fig:LHC14TeV}}
\end{figure}

We perform the numerical calculation of the lowest order cross section 
for the process $pp \rightarrow Z' \rightarrow \ttbar$ at the LHC 
for different values of $\sqrt{s}$ between $7-14$~TeV, using 
Eq.~\ref{eq_tot_xsec} and taking into account the 
spin-color factor in Eq.~\ref{eq_spin_color}. 
We calculate the cross section for different choices of 
$\Gamma_{Z'}$, equal to $1\%$, $1.2\%$, $2\%$, and $10\%$ of $M_{Z'}$.
The first three widths qualify as narrow resonances at
the LHC, and the integration in Eq.~\ref{eq_tot_xsec} is
performed using the full available phase space of 
$ 2m_t < m < \sqrt{s}$. The integration for 
$\Gamma_{Z'} = 10\%M_{Z'}$ is performed using the mass interval 
$M_{Z'} - 3\Gamma_{Z'} < m < M_{Z'} + 3\Gamma_{Z'}$ in order to sample
better the cross section around the peak of the resonance. 
The results are tabulated in 
Tables~\ref{tab:ZprimexsecLHC7}-\ref{tab:ZprimexsecLHC14}, 
and displayed in Figs.~\ref{fig:LHC7TeV}-\ref{fig:LHC14TeV}. 
%
%
%
\section{Conclusions}
\label{sec:conclusions}
We have presented cross section calculations of the leptophobic 
topcolor $Z'$ decaying to $\ttbar$. These calculations update the 
results presented in Ref.~\cite{hep-ph-9911288} for the Tevatron, by
both fixing an error in the reported width of the leptophobic topcolor 
$Z'$, and using $m_t=172.5$ GeV/c$^2$ and CTEQ6L parton 
distributions in an improved calculation procedure. This 
note documents the first calculations of the cross section 
for a leptophobic topcolor $Z'$ at the LHC.

\section{Acknowledgments}
\label{sec:acknowledgments}
We thank Chris Hill for sending 
us the correction to the equation for the 
topcolor $Z'$ width, found by James Ferrando 
and Mads Frandsen. We also thank 
Jan Steggemann and Ia Iashvili for several 
useful discussions.


%
\begin{table*}[htbH]
\begin{center}
\caption{Cross sections at the LHC at $\sqrt{s} = 7$~TeV for 
$\sigma_{Z'}{\rm B(}Z'\rightarrow\ttbar{\rm )}$ at different $Z'$ 
masses and widths.
\label{tab:ZprimexsecLHC7}}
\begin{ruledtabular}
\begin{tabular}{ccccc}
$M_{Z'}$ & \multicolumn{4}{c}{$\sigma_{Z'}{\rm B(}Z'\rightarrow\ttbar{\rm )}$
  [pb]} \\ \cline{2-5}
[GeV/c$^2$]
& $\Gamma_{Z'}/M_{Z'} = 0.01$ 
& $\Gamma_{Z'}/M_{Z'} = 0.012$  
& $\Gamma_{Z'}/M_{Z'} = 0.02$  
& $\Gamma_{Z'}/M_{Z'} = 0.10$ \\
\hline 
400.0	&	24.82	&	29.61	&	48.24	&	220.92	\\
500.0	&	12.17	&	14.61	&	24.37	&	115.29	\\
600.0	&	6.76	&	8.07	&	13.27	&	60.37	\\
700.0	&	3.66	&	4.39	&	7.29	&	33.07	\\
750.0	&	2.74	&	3.28	&	5.45	&	24.95	\\
800.0	&	2.05	&	2.47	&	4.13	&	19.00	\\
900.0	&	1.20	&	1.44	&	2.43	&	11.27	\\
1000.0	&	0.753	&	0.905	&	1.51	&	6.91	\\
1250.0	&	0.236	&	0.283	&	0.477	&	2.21	\\
1500.0	&	8.20E-02	&	9.87E-02	&	0.167	&	0.777	\\
1700.0	&	3.71E-02	&	4.48E-02	&	7.67E-02	&	0.352	\\
2000.0	&	1.16E-02	&	1.41E-02	&	2.48E-02	&	0.113	\\
2500.0	&	1.90E-03	&	2.36E-03	&	4.42E-03	&	1.81E-02	\\
3000.0	&	3.59E-04	&	4.64E-04	&	9.95E-04	&	3.01E-03	\\
\end{tabular}
\end{ruledtabular}
\end{center}
\end{table*}
\begin{table*}[htbH]
\begin{center}
\caption{Cross sections at the LHC at $\sqrt{s} = 8$~TeV for 
$\sigma_{Z'}{\rm B(}Z'\rightarrow\ttbar{\rm )}$ at different $Z'$ 
masses and widths.
\label{tab:ZprimexsecLHC8}}
\begin{ruledtabular}
\begin{tabular}{ccccc}
$M_{Z'}$ & \multicolumn{4}{c}{$\sigma_{Z'}{\rm B(}Z'\rightarrow\ttbar{\rm )}$
  [pb]} \\ \cline{2-5}
[GeV/c$^2$]
& $\Gamma_{Z'}/M_{Z'} = 0.01$ 
& $\Gamma_{Z'}/M_{Z'} = 0.012$  
& $\Gamma_{Z'}/M_{Z'} = 0.02$  
& $\Gamma_{Z'}/M_{Z'} = 0.10$ \\
\hline
400.0	&	29.97	&	35.77	&	58.46	&	271.98	\\
500.0	&	14.79	&	17.82	&	29.98	&	145.06	\\
600.0	&	8.42	&	10.08	&	16.70	&	77.71	\\
700.0	&	4.78	&	5.74	&	9.57	&	43.59	\\
750.0	&	3.59	&	4.31	&	7.17	&	33.29	\\
800.0	&	2.76	&	3.32	&	5.54	&	25.67	\\
900.0	&	1.64	&	1.98	&	3.35	&	15.63	\\
1000.0	&	1.03	&	1.24	&	2.08	&	9.84	\\
1250.0	&	0.367	&	0.441	&	0.742	&	3.37	\\
1500.0	&	0.133	&	0.160	&	0.273	&	1.28	\\
1700.0	&	6.62E-02	&	7.99E-02	&	0.136	&	0.616	\\
2000.0	&	2.26E-02	&	2.75E-02	&	4.75E-02	&	0.218	\\
2500.0	&	4.33E-03	&	5.30E-03	&	9.52E-03	&	4.21E-02	\\
3000.0	&	9.30E-04	&	1.16E-03	&	2.26E-03	&	8.59E-03	\\
\end{tabular}
\end{ruledtabular}
\end{center}
\end{table*}
\begin{table*}[htbH]
\begin{center}
\caption{Cross sections at the LHC at $\sqrt{s} = 10$~TeV for 
$\sigma_{Z'}{\rm B(}Z'\rightarrow\ttbar{\rm )}$ at different $Z'$ 
masses and widths.
\label{tab:ZprimexsecLHC10}}
\begin{ruledtabular}
\begin{tabular}{ccccc}
$M_{Z'}$ & \multicolumn{4}{c}{$\sigma_{Z'}{\rm B(}Z'\rightarrow\ttbar{\rm )}$
  [pb]} \\ \cline{2-5}
[GeV/c$^2$]
& $\Gamma_{Z'}/M_{Z'} = 0.01$ 
& $\Gamma_{Z'}/M_{Z'} = 0.012$  
& $\Gamma_{Z'}/M_{Z'} = 0.02$  
& $\Gamma_{Z'}/M_{Z'} = 0.10$ \\
\hline
400.0	&	39.51	&	47.28	&	77.96	&	377.58	\\
500.0	&	22.69	&	27.22	&	45.29	&	207.45	\\
600.0	&	12.08	&	14.51	&	24.20	&	114.59	\\
700.0	&	7.23	&	8.64	&	14.35	&	66.36	\\
750.0	&	5.53	&	6.66	&	11.14	&	51.51	\\
800.0	&	4.50	&	5.38	&	8.91	&	40.38	\\
900.0	&	2.80	&	3.37	&	5.65	&	25.44	\\
1000.0	&	1.79	&	2.15	&	3.61	&	16.59	\\
1250.0	&	0.673	&	0.811	&	1.36	&	6.24	\\
1500.0	&	0.283	&	0.340	&	0.571	&	2.61	\\
1700.0	&	0.145	&	0.175	&	0.298	&	1.37	\\
2000.0	&	5.92E-02	&	7.12E-02	&	0.121	&	0.552	\\
2500.0	&	1.44E-02	&	1.74E-02	&	2.99E-02	&	0.135	\\
3000.0	&	3.74E-03	&	4.57E-03	&	8.16E-03	&	3.57E-02	\\
4000.0	&	3.33E-04	&	4.24E-04	&	8.67E-04	&	2.79E-03	\\
\end{tabular}
\end{ruledtabular}
\end{center}
\end{table*}
\begin{table*}[htbH]
\begin{center}
\caption{Cross sections at the LHC at $\sqrt{s} = 14$~TeV for 
$\sigma_{Z'}{\rm B(}Z'\rightarrow\ttbar{\rm )}$ at different $Z'$ 
masses and widths.
\label{tab:ZprimexsecLHC14}}
\begin{ruledtabular}
\begin{tabular}{ccccc}
$M_{Z'}$ & \multicolumn{4}{c}{$\sigma_{Z'}{\rm B(}Z'\rightarrow\ttbar{\rm )}$
  [pb]} \\ \cline{2-5}
[GeV/c$^2$]
& $\Gamma_{Z'}/M_{Z'} = 0.01$ 
& $\Gamma_{Z'}/M_{Z'} = 0.012$  
& $\Gamma_{Z'}/M_{Z'} = 0.02$  
& $\Gamma_{Z'}/M_{Z'} = 0.10$ \\
\hline 
400.0	&	58.83	&	70.86	&	119.29	&	597.95	\\
500.0	&	37.60	&	45.08	&	74.65	&	339.69	\\
600.0	&	21.05	&	25.43	&	42.73	&	194.07	\\
700.0	&	13.12	&	15.66	&	25.69	&	116.34	\\
750.0	&	9.80	&	11.81	&	19.80	&	91.89	\\
800.0	&	8.05	&	9.68	&	16.19	&	73.33	\\
900.0	&	5.12	&	6.16	&	10.31	&	47.91	\\
1000.0	&	3.48	&	4.19	&	7.07	&	32.44	\\
1250.0	&	1.53	&	1.83	&	3.03	&	13.47	\\
1500.0	&	0.675	&	0.814	&	1.37	&	6.24	\\
1700.0	&	0.393	&	0.471	&	0.784	&	3.57	\\
2000.0	&	0.178	&	0.214	&	0.360	&	1.65	\\
2500.0	&	5.44E-02	&	6.57E-02	&	0.112	&	0.515	\\
3000.0	&	1.93E-02	&	2.32E-02	&	3.95E-02	&	0.178	\\
4000.0	&	2.66E-03	&	3.24E-03	&	5.76E-03	&	2.50E-02	\\
\end{tabular}
\end{ruledtabular}
\end{center}
\end{table*}

\end{document}